\documentclass[]{spie}  

\usepackage{amsmath,amsfonts,amssymb}
\usepackage{graphicx}
\usepackage[colorlinks=true, allcolors=blue]{hyperref}

\title{Achieving a Spectropolarimetric Precision Better than 0.1\% in the Near-Infrared with WIRC+Pol}

\author[a]{Samaporn Tinyanont}
\author[a,b,e]{Maxwell Millar-Blanchaer}
\author[a]{Nemanja Jovanovic}
\author[a,b]{Dimitri Mawet}
\author[b]{Gautam Vasisht}
\author[a]{Jennifer W. Milburn}
\author[b]{Eugene Serabyn}
\author[a]{Michael Porter}
\author[c]{Skyler Palatnick}
\author[d]{Connor Hopkins}
\affil[a]{Department of Astronomy, California Institute of Technology, 1200 E California Blvd, MC 249-17, Pasadena, CA 91125, USA}
\affil[b]{Jet Propulsion Laboratory, California Institute of Technology, 4800 Oak Grove Dr, Pasadena, CA 91109, USA}
\affil[c]{Department of Physics and Astronomy, University of Pennsylvania, 209 S 33rd St, Philadelphia, PA 19104}
\affil[d]{Pasadena High School, 2925 E Sierra Madre Blvd Pasadena, CA 91107}
\affil[e]{NASA Hubble Fellow}

\authorinfo{Further author information: (Send correspondence to S.T.)\\S.T.: E-mail: st@astro.caltech.edu}

\begin{document} 
\maketitle

\begin{abstract}
WIRC+Pol is a near-infrared low-resolution spectropolarimeter on the 200-inch Telescope at Palomar Observatory.
The instrument utilizes a polarization grating to perform polarimetric beam splitting and spectral dispersion simultaneously.
It can operate either with a focal plane slit to reduce sky background or in a slitless mode. 
Four different spectra sampling four linear polarization angles are recorded in the focal plane, allowing the instrument to measure all linear polarization states in one exposure.
The instrument has been on-sky since February 2017 and we found that the systematic errors, likely arising from flat fielding and gravity effects on the instrument, limit our accuracy to $\sim$1\%.
These systematic effects were slowly varying, and hence could be removed with a polarimetric modulator.
A half-wave plate modulator and a linear polarizer were installed in front of WIRC+Pol in March 2019. 
The modulator worked as expected, allowing us to measure and remove all instrumental polarization we previously observed.
The deepest integration on a bright point source ($J$ = 7.689, unpolarized star HD\,65970) demonstrated uncertainties in $q$ and $u$ of 0.03\% per spectral channel, consistent with the photon noise limit. 
Observations of fainter sources showed that the instrument could reach the photon noise limit for observations in the slitless mode.
For observations in slit, the uncertainties were still a factor of few above the photon noise limit, likely due to slit loss.


\end{abstract}


\section{INTRODUCTION}
Polarization is a fundamental property of light that is often left unmeasured by astronomical observations largely due to its difficulties and the lack of dedicated instrumentation. 
The greatest challenge for measuring polarization for astronomical sources is that, in many cases, the polarization signal of interest is much smaller than 1\%, requiring a large number of photons to establish a significant detection: $10^6$ photons for 0.1\% precision.
The problem is exacerbated in the infrared (IR) where the bright sky background further limits the sensitivity. 
Current IR polarimeters offer either no spectral information but good sensitivity for faint sources or medium resolving power ($R\approx1000$) but only sensitive to bright sources.
The examples of these instruments are the Long-slit Intermediate Resolution Infrared Spectrograph (LIRIS) on the William Herschel Telescope\cite{manchado2004} and the Infrared Camera and Spectrograph (IRCS) on Subaru Telescope\cite{watanabe2018, tokunaga1998}.
Another IR spectropolarimeter available is SPectropolarimètre InfraROUge (SPIRou)\cite{donati2018} on Canada-France-Hawaii Telescope (CFHT) on Maunakea, which operates at high spectral resolution ($R\approx 75000$).
The high resolution combined with the smaller telescope limits SPIRou to observe bright stellar targets. 

\begin{figure}
    \centering
    \includegraphics[width = 0.7\textwidth]{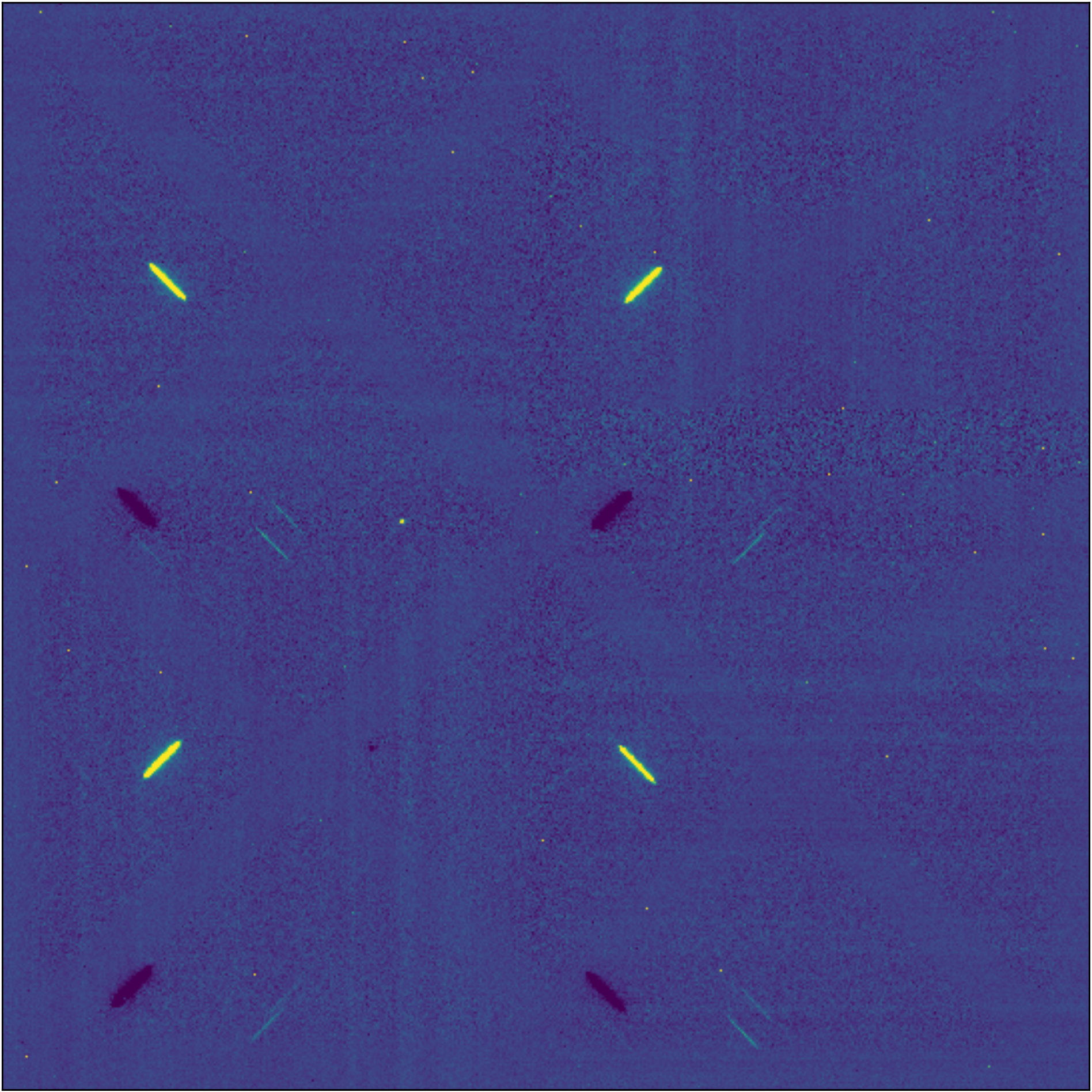}
    \caption{WIRC+Pol background subtracted data with a bright standard star in the field of view.
    The four spectral traces probe four orthogonal angles of linear polarization.
    At the center of the four traces is the undispersed, zeroth order light of the star. 
    Faint traces of a few background stars in the field of view could be seen. }
    \label{fig:data}
\end{figure}

In February 2017, we installed a polarization grating (PG\cite{millar2014, packham2010, escuti2006}) into the Wide-field InfraRed Camera (WIRC\cite{wilson2003}) at the prime focus of the 200-inch Hale Telescope at Palomar Observatory (P200).
The resulting low-resolution ($R\approx 100$) near-IR (1.1-1.8 $\mu$m, \textit{JH} bands) spectropolarimeter was named WIRC+Pol, and its design, data reduction pipeline, and instrumental characteristics were discussed in details in Ref.~\citenum{tinyanont2019}. 
The original instrument was designed such that there were four spectral traces on the detector for one source; each of the traces correspond to 0$^\circ$, 90$^\circ$, 45$^\circ$, and 135$^\circ$ angles of linear polarization.
As such, linear polarization could be measured in one shot by comparing fluxes in these four spectral traces in the different quadrants of the detector. 
Fig.~\ref{fig:data} shows WIRC+Pol data with one source in the field of view. 

During the commissioning of the instrument, we uncovered that the instrumental systematics were strong, on the order of a few percent, and was highly dependent on the telescope pointing, as shown in Fig.~6 and 8 in Ref.~\citenum{tinyanont2019}.
The Hale telescope's polarization had been measured to be below 10 ppm\cite{wiktorowicz2009}, so the systematics observed were from the instrument.
This limited the instrument's accuracy, as it required an unpolarized standard star to be observed at a telescope pointing no further than about 10$^\circ$ from the source to obtain a systematic correction better than $\sim$1\%. 
Moreover, the filter transmission profiles seen by the four traces were not the same.
This was because the PG was installed upstream from the tilted broadband interference filter, resulting in the spectral traces entering the filter at different incident angles. 
The shifted transmission profile made wavelength alignment between the four traces difficult, and we had to rely on precisely aligning an atmospheric absorption feature in the spectra.
Even then, the resulting polarization measurements still had significant spurious wavelength dependent systematic in them (e.g. Fig.~6 left in Ref.~\citenum{tinyanont2019}).
It is crucial to note that these systematics only vary slowly with time (Fig.~8 in Ref.~\citenum{tinyanont2019}), and could be removed using a polarimetric modulator. 
Lastly, the focal plane mask of the instrument was designed to have an open area for slitless spectropolarimetry of bright sources, and three 3" circular holes to act as slits to limit sky background for faint sources. 
The circular design was chosen for two considerations: to reduce potential slit-induced polarization and to make it easier to manufacture on a traditional stainless steel substrate.
We learned during commissioning that the circular slits required pointing and guiding precision beyond the telescope and instrument's capabilities, owing largely to the fact that P200 does not have a separate guider for prime focus instruments, including WIRC.

In this work, we briefly describe the installation of a half-wave plate (HWP) modulator and a linear polarizer in front of WIRC+Pol, done in early 2019, and we will focus on the instrumental characterization post-upgrade. 
We will also discuss the novel slit design used to make a physically small rectangular slit required for the prime focus instrument. 
\textsection\ref{sec:hwp} describes the upgrade and how the HWP allows us to measure and remove the systematics and \textsection\ref{sec:slit} focuses on the new slit design. 
We explain our current observation technique with the HWP in \textsection\ref{sec:obs}. 
We show the instrumental characterization result, illustrating that the systematics have been removed for sources bright and faint in \textsection\ref{sec:remove_ip}. 
We discuss the future characterization of the polarimetric efficiency and cross-talk, i.e. constructing the instrumental Mueller matrix for WIRC+Pol, using observations of polarized stars and observations of unpolarized stars with the linear polarizer in \textsection\ref{sec:mueller}.
Finally, we present our conclusions in \textsection\ref{sec:conclusion}.

\section{Half-wave plate and linear polarizer upgrades}\label{sec:hwp}
The key hardware upgrade used to address the instrumental systematics from WIRC+Pol was the installation of a HWP and a LP in front of WIRC+Pol.
WIRC+Pol only operated in the $J$ and $H$ bands and not in the thermal IR, which allowed the optics to be installed outside of the cryogenic dewar to reduce cost.
Otherwise (in the $K$ band and beyond), the IR emission from warm optics would contribute significantly to the background.
The location of the optics in the f/3.3 converging beam presented strict requirements on the optics.
The details of the optomechanical design of this upgrade will be presented in a future publication.
Briefly, we installed a HWP and a LP on linear and rotation stages in front of WIRC, so that they could be inserted and removed from the beam, and rotated to an arbitrary angles independently.
The optics sizes and placement were such that the 4.3' polarimetric field of view was not vignetted. 
The positions and rotation angles of the HWP and LP are stored in the header of images taken. 
The LP converts light from any source into 100\% polarized light, which can be used to measure the polarimetric efficiency and cross-talk terms in the Mueller matrix of WIRC+Pol. 
This characterization work is deferred also for a future publication. 

\subsection{Polarization calculation: flux ratio method}
\begin{figure}
    \centering
    \includegraphics[width=\textwidth]{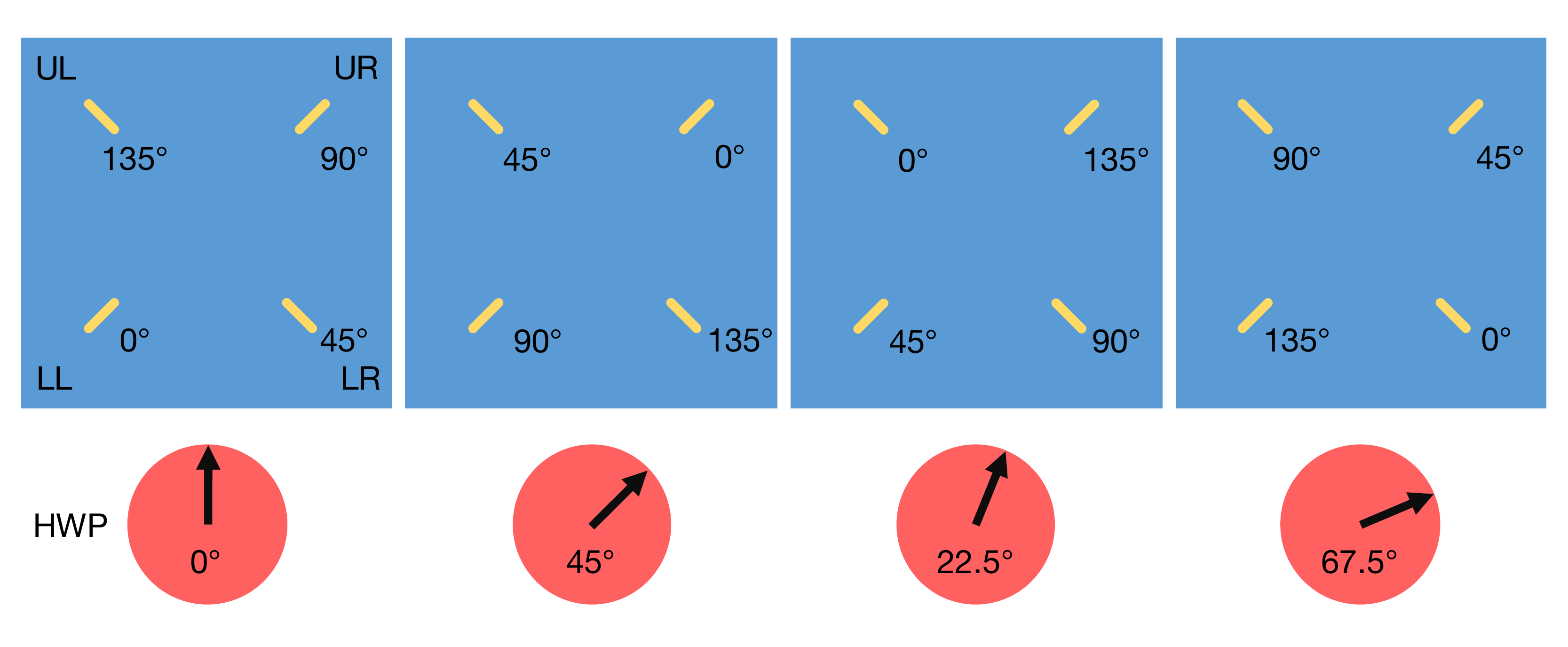}
    \caption{A schematic showing the four HWP angles used in a typical observing sequence: 0$^\circ$, 45$^\circ$, 22.5$^\circ$, and 67.5$^\circ$, and the corresponding linear polarization angle probed by the four spectral traces. 
    At 0$^\circ$, the linear polarization angle sampled by each of the four traces (Upper Left, Lower Right, Upper Right, and Lower Left) are as indicated on the figure. These angles were determined by observing twilight polarization (\textsection 4.2 in Ref.~\citenum{tinyanont2019}).   
    }
    \label{fig:hwp}
\end{figure}

We now demonstrate that calculating the normalized Stokes parameters using the observations at two different HWP angles effectively removes the systematic errors old WIRC+Pol suffered.
We first note that a rotation of the HWP by $\theta$ rotates the incoming linear polarization angle by $2\theta$. 
As a result, by rotating the HWP to 0$^\circ$, 45$^\circ$, 22.5$^\circ$, and 67.5$^\circ$, we can cycle all four spectral traces through four different linear polarization angles, as visualized in Fig.~\ref{fig:hwp}.
For simplicity, we consider $q$ computed from the lower left (LL) and upper right (UR) traces, which, without the HWP, sample 0$^\circ$ and 45$^\circ$ respectively. 
For a source with intrinsic polarization $q$ and the total flux $I$, the observed fluxes at the HWP angle of 0$^\circ$ are
\begin{equation}
    S_{\rm LL,0} (\lambda) = (1+q) I (\lambda) F_{\rm LL} (\lambda) A_{t_1}(\lambda)  \hspace{1in}  S_{\rm UR,0} (\lambda) = (1-q) I (\lambda) F_{\rm UR} (\lambda) A_{t_1}(\lambda) 
\end{equation}
where $F_{i}$ is the filter transmission profile seen by the $i$ trace and $A_{t_1}$ is the atmospheric transmissions at time $t_1$. 
At the orthogonal HWP angle of 45$^\circ$ observed at a later time $t_2$, the angle of polarization probed by the two traces flips: 
\begin{equation}
    S_{\rm LL,45} (\lambda) = (1-q) I (\lambda) F_{\rm LL} (\lambda) A_{t_2}(\lambda)  \hspace{1in}  S_{\rm UR,45} (\lambda) = (1+q) I (\lambda) F_{\rm UR} (\lambda) A_{t_2}(\lambda) 
\end{equation}
In order to recover the intrinsic polarization $q$ from these observations, we employ the flux ratio method presented in Ref.~\citenum{zapatero2011}, eq. (3) and (4).
Following that formulation, we define a quantity
\begin{align}
    R_q^2 &= \dfrac{S_{\rm LL,0}/S_{\rm UR, 0}}{S_{\rm LL,45}/S_{\rm UR, 45}} \\ 
    &= \dfrac{(1+q)F_{\rm LL}/(1-q)F_{\rm UR}}{(1-q)F_{\rm LL}/(1+q)F_{\rm UR}} \\
    &= \dfrac{(1+q)^2}{(1-q)^2}
\end{align}
The uncertainties of this quantity is simply a quadrature sum:
\begin{equation}
    d R_q^2 = R_q^2 \sqrt{ \left(\dfrac{dS_{\rm LL,0}}{S_{\rm LL,0}}\right)^2 +
                            \left(\dfrac{dS_{\rm LL,45}}{S_{\rm LL,45}}\right)^2+
                            \left(\dfrac{dS_{\rm UR,0}}{S_{\rm UR,0}}\right)^2+
                            \left(\dfrac{dS_{\rm UR,45}}{S_{\rm UR,45}}\right)^2}
\end{equation}
Then, $q$ could be recovered by computing
\begin{equation}
       \dfrac{R_q - 1}{R_q + 1} = q
\end{equation}
The polarimetric uncertainty is 
\begin{equation}
    dq = \dfrac{dR_q^2}{(R_q +1 )^2 R_q^2}
\end{equation}
Similar calculation can be done to obtain $u$. 

\subsection{Limitations of the double difference method}
Another primary method of computing polarization with a dual-beam polarimeter is to perform double difference.
This is done by computing, for instance, $q$ before and after modulating the HWP, then averaging the two. 
For WIRC+Pol, this would be
\begin{align}
    q_1 &= \dfrac{S_{\rm LL, 0}  -  S_{\rm UR, 0}}{S_{\rm LL, 0}  +  S_{\rm UR, 0}} \\
    q_2 &= \dfrac{S_{\rm UR, 45} -  S_{\rm LL,  45}}{S_{\rm UR, 45} -  S_{\rm LL, 45}}
\end{align}
and the double differenced polarization is $q_{\rm DD} = (q_1 + q_2)/2$. 
Writing down the full expressions for $S_{i,j}$, the previous quantity becomes
\begin{align}
    2q_{\rm DD} &= q_1 + q_2 \\
                &=  \dfrac{ (1+q) F_{\rm LL} - (1-q) F_{\rm UR}}{(1+q) F_{\rm LL} + (1-q) F_{\rm UR}} + \dfrac{ (1+q) F_{\rm UR} - (1-q) F_{\rm LL}}{(1+q) F_{\rm UR} + (1-q) F_{\rm LL}}
\end{align}
This does not generally simplify to $q_{\rm DD} = q$ due to the filter transmission difference between the four traces.
This is also true for non-common path effects between the ordinary and extraordinary beams in a dual-beam polarimeter in general, and not just the filter transmission issue in our instrument.

One can alleviate this by computing the differencing first using the same trace at two HWP angles:
\begin{align}
    q_1 &= \dfrac{S_{\rm LL, 0}  -  S_{\rm LL, 45}}{S_{\rm LL, 0}  +  S_{\rm LL, 45}} \\
    q_2 &= \dfrac{S_{\rm UR, 45} -  S_{\rm UR,  0}}{S_{\rm UR, 45} -  S_{\rm UR, 0}}
\end{align}
and the double differenced polarization is, instead, $q_{\rm DD} = (q_1 + q_2)/2$. 
Writing down the full expressions for $S_{i,j}$, the previous quantity becomes
\begin{align}
    2q_{\rm DD} &= q_1 + q_2 \\
                &=  \dfrac{ (1+q) A_{t_1} - (1-q) A_{t_2}}{(1+q) A_{t_1} + (1-q) A_{t_2}} + \dfrac{ (1+q) A_{t_2} - (1-q) A_{t_1}}{(1+q) A_{t_2} + (1-q) A_{t_1}}
\end{align}
This formulation works better.
It could simplify to $q_{\rm DD} = q$ if the atmospheric effects did not change between the two observations, i.e. $A_{t_1} = A_{t_2}$. 
Furthermore, both double differencing methods work for an unpolarized source $q = 0$.
In summary, the flux ratio is more robust against non-common path effects and atmospheric changes during observations.

\section{New focal plane mask with slit}\label{sec:slit}
\subsection{Issues with old circular slits}
The focal plane mask of WIRC+Pol was also replaced in the 2019 upgrade. 
The first version of the mask was fabricated using the wire EDM process on stainless steel substrate. 
The shape of the slit mask could be seen in Fig.~2 (center) in Ref.~\citenum{tinyanont2019}. 
In the center of the mask, there were three holes, 245 $\mu$m in diameter, corresponding to 3" on sky at the f/3.3 focal plane of a 5.1-meter primary mirror.
The decision to incorporate circular slits instead of a traditional rectangular slit was two-fold. 
First, circular slits in principle would reduce the amount of slit-induced polarization, as the symmetric shape of the slit would make the polarization created by the metal edge of the slit cancel out.
Second, at the focal length at the prime focus, the physical size of the slit would be minute, making the fabrication difficult on aluminum or stainless steel.

From the first two years of operation with WIRC+Pol, we learned that the slit holes presented challenges for telescope control, largely due to WIRC's lack of a separate guider camera to assist with pointing and guiding. 
First, the target acquisition was time consuming, and a slight offset from the center of the hole would result in large slit loss, which diminished the benefit gained from reduced sky background. 
Further, once the target was acquired, it was difficult to guide the telescope to keep the target inside the slit. 
This challenge was alleviated in cases where there were nearby sources in the slitless area of the focal plane, as we could guide on them instead. 
Lastly, it was difficult to dither the telescope to shuffle the source between three holes in order to perform sky subtraction, as there were no guider images to confirm the pointing. 
As a result, the slit holes remained largely unused, limiting WIRC+Pol's observations to relatively bright sources in the higher background slitless area. 

\subsection{New slit design}
For the instrument upgrade, we also planned to install a new mask with a rectangular slit, in order to ease the operational requirements. 
The rectangular slit would retain the width of 3" to minimize slit loss provided the median seeing at Palomar of 1.5". 
The length of the slit was designed such that we would be able to observe a source in 2" seeing condition, in two dither positions in the slit that were 10$\sigma$ apart, where $\sigma$ is the standard deviation of the Gaussian (seeing limited) PSF of the source. 
This was to ensure that the source could be put into two different positions inside the slit so that an image observed with source at position A could use another image with source at position B as a background image for subtraction without the source self-subtracting.
The slit length required to do this was 17".
Thus, the physical size of the slit would be 245 $\times$ 1388 $\mu$m.
The small physical size of the slit was difficult to archive precisely with the wire EDM fabrication used for the old slit mask.
Instead, the new slit was manufactured on a silicon substrate using the electron-beam lithography (EBL) technique, which allowed for the slit to be cut into the substrate with high precision.
The small rectangular piece of silicon with slit was then held in place by an aluminum mask (Fig.~\ref{fig:slit} top right).

\subsection{Slit-induced polarization}
The new rectangular slit would lack the symmetry of the old circular slit holes, and might induce polarization in the observed source. 
Calculation showed, however, that the expected slit-induced polarization should be very low.
The slit-induced polarization is a function of slit width, depth (thickness of the slit in the direction of light propagation), and the conductivity of the slit material. 
We estimated the slit-induced polarization using the relations derived by Ref.~\citenum{slater1959}. 
\begin{align}
a_s &= \dfrac{2}{b} \sqrt{\dfrac{\omega \epsilon_0}{2 \sigma}}
        \left( 1- \left(\dfrac{n \lambda}{2 b} \right)^2 \right)^{-1/2} \\
a_p &= a_s \left(\dfrac{n \lambda}{2 b} \right)^2 \\
    p_{\rm slit} &= \dfrac{e^{-2 a_p z} -  e^{-2 a_s z}}{e^{-2 a_p z} +  e^{-2 a_s z}} 
\end{align}
where $b$ is the slit width, $z$ is thickness, $\sigma$ is the conductivity of the material, $\omega$ is the angular frequency of light, and $\lambda$ is the wavelength.
Parameters are in SI units, and $\epsilon_0$ is the permittivity of free space.
With the wire-EDM aluminum mask, the minimum thickness was 100 $\mu$m, resulting in the expected slit polarization of order 1\%.
In the original design, this high potential of slit induced polarization informed the decision to have circular holes to try to mitigate it. 
However, for the EBL silicon slit, the edge of the slit could be made to be only 9 $\mu$m thick.
While the surface of the silicon slit was coated with metal to make it opaque, the bulk of the material was not a conductor, further reducing the expected slit-induced polarization. 
The upper limit of the slit-induced polarization from this design was 0.1\%, calculated for aluminum's conductivity. 
(Equations above are only valid for conductors. 
For dielectric materials like silicon, Ref.~\citenum{ismail1986} showed that slit-induced polarization is lower.)
Therefore, we concluded that the rectangular shape of the new slit would not create appreciable slit-induced polarization. 
We will show with data in \textsection\ref{sec:remove_ip} that this is the case. 

\subsection{Slit dimensions measurement}
\begin{figure}
    \centering
    \includegraphics[width=\textwidth]{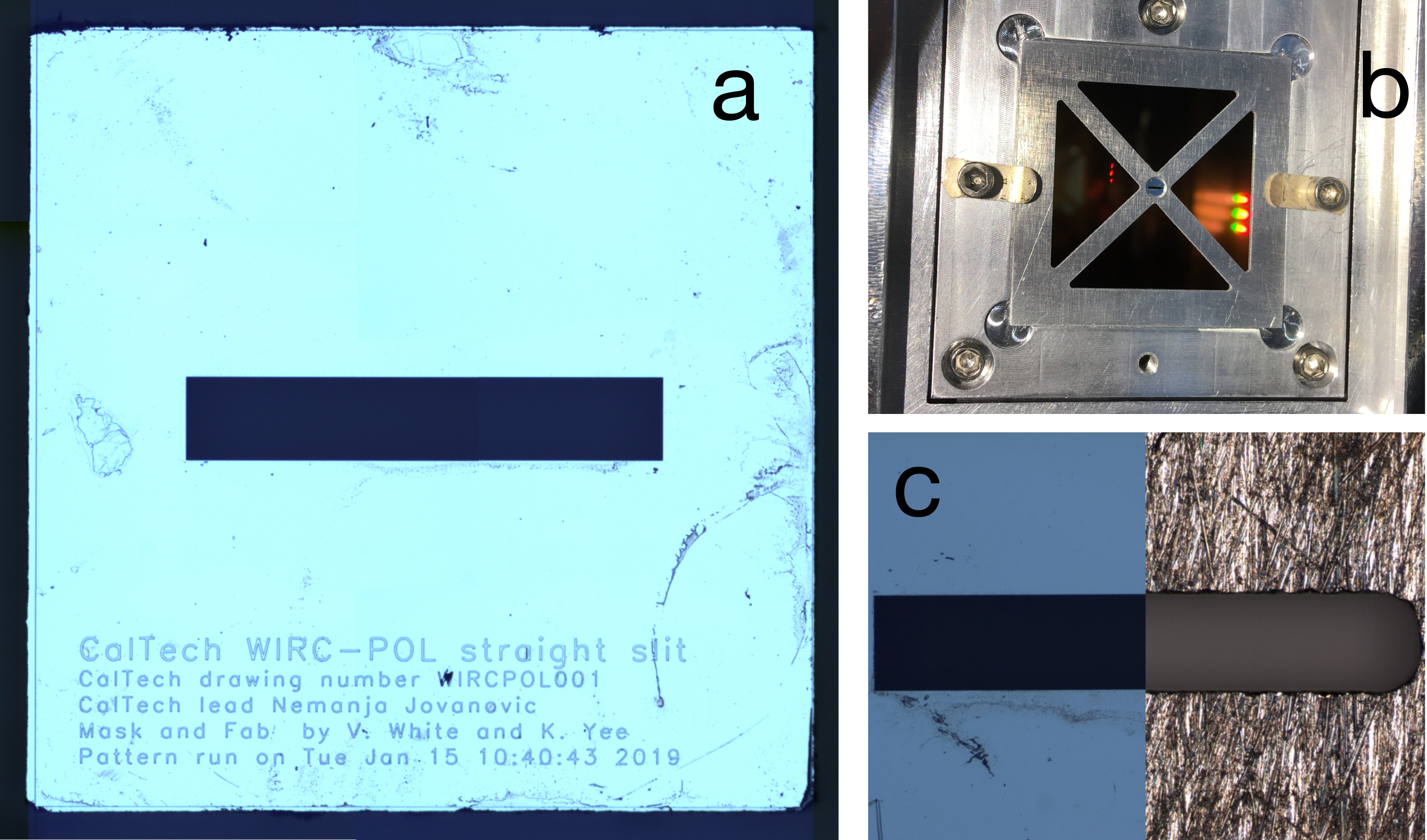}
    \caption{Left (a): image of the full silicon piece with the slit in the center. 
    The dimensions were measured 241 by 1380 $\mu$m, within a few microns from designed dimensions. 
    Top Right (b): the aluminum coated side of the slit installed in the new aluminum focal plane mask, both held in place in the focal plane mask mechanism inside WIRC.
    Bottom Right (c): the comparison between the silicon slit and the aluminum slit manufactured to the same specifications.}
    \label{fig:slit}
\end{figure}

After the fabrication, we measured the dimensions of the slit using microscope images, and found that they were within a few microns of the specifications at 241.5 $\times$ 1380 $\mu$m.
Fig.~\ref{fig:slit} (left) shows the image of the full slit that was installed in WIRC+Pol, along with (bottom right) a comparison between it and a reference aluminum slit of the same designed dimension. 
Note the vast improvement in the roughness of the edges, and how sharp the corners were on the silicon slit in comparison to the aluminum slit.
The slit width is uniform, with 0.1 $\mu$m variation, across the length of the slit.
Fig.~\ref{fig:slit} (top right) shows the silicon slit piece held in place in the new aluminum focal plane mask, which was installed in the focal mask mechanism inside WIRC.

\section{Observational technique}\label{sec:obs}
\subsection{Modulation sequence}
Along with the upgrade, the instrument control software was updated to take simple command scripts in order to automatize observing sequences.
In order to remove the slowly varying systematics, we observed a source for up to 2 min in one HWP angle (this may include several individual frames or coadds), then we rotated the HWP such that the angle of polarization is rotated to an orthogonal angle (0$^\circ$ to 90$^\circ$, and 45$^\circ$ to 135$^\circ$). 
The 1-2 min limit were informed by the timescale at which the systematics changed.
The overhead from the HWP rotation was minimal by design, approximately a second, which could be timed to coincide with detector readout in a future software upgrade.
The systematics in WIRC+Pol prior to the HWP upgrade were not only a function of time, but also a strong function of the detector position of the source, with deviations of only 30" resulting in a 1\% change in the systematics (Fig.~7 and \textsection4.5 in Ref.~\citenum{tinyanont2019}). 
With the HWP, we were able to remove the systematics for any given detector location.
As a result, the strict guiding requirements for the old WIRC+Pol in order to assure that sources and standard stars were observed at the same detector position no longer applied. 
As we will show in \textsection\ref{sec:remove_ip}, the HWP modulator eliminated the need to observe an unpolarized standard star immediately before or after a science target to measure the instrumental polarization systematics, since the modulator allowed us to measure and remove those systematics for any source.
Data also showed that the telescope polarization, and other instrumental polarization upstream of the waveplate was minimal.
Observations have shown, however, that one might still need to observe either a polarized standard or an unpolarized standard with a linear polarizer to measure the angle of polarization zero point, and other Mueller matrix terms (polarization efficiency and cross-talk).

\subsection{In-slit observation}
For faint objects ($J \gtrsim 13$), objects in crowded fields, and extended sources, the observations are best performed with the source inside the slit to minimize sky background. 
The trade-off is that the sky background, while lower, is no longer spatially smooth as sky lines and focal plane mask features can be seen around the source trace. 
This requires image subtraction in order to remove the structured background.
In order to do this, we observe a source in slit at a position ``A" on one side of the slit at four HWP angles, then dither the telescope to move the source to the ``B" position where the source is observed also at four HWP angles in the same rotation sequence. 
Due to the lack of a guiding camera, it is occasionally necessary to remove the PG and take a direct image of the slit to ensure that the source is sill well centered. 
The images at the ``B" position are then used as background images to be subtracted from images at the ``A" positions at the same HWP angle.
The WIRC+Pol data reduction pipeline (DRP), described in Ref.~\citenum{tinyanont2019} and publicly accessible at \url{https://github.com/WIRC-Pol/wirc_drp}, has been updated to reduce the new data with HWP.
The pipeline applies the calculation described in \textsection\ref{sec:hwp} to compute double differencing $q$ and $u$ for each of the sets of four images with four HWP angles.

\section{Calibrating instrumental polarization}\label{sec:remove_ip}
In this section, we will demonstrate the instrument's performance post HWP upgrade, to show that the instrumental systematics have been effectively removed from WIRC+Pol. 
On the first night after the HWP installation, 2019 March 17 UT, we observed several unpolarized standard stars using the observation techniques described above. 
The list of unpolarized stars was obtained from Ref.~\citenum{heiles2000}. 
For these bright stars, the observations were obtained outside of the slit. 
The data were processed using the WIRC+Pol DRP, which applied dark subtraction and flat fielding, spectral extraction, and polarization calculation.

\subsection{Bright unpolarized standard star}
\begin{figure}
    \centering
    \includegraphics[width=0.8\textwidth]{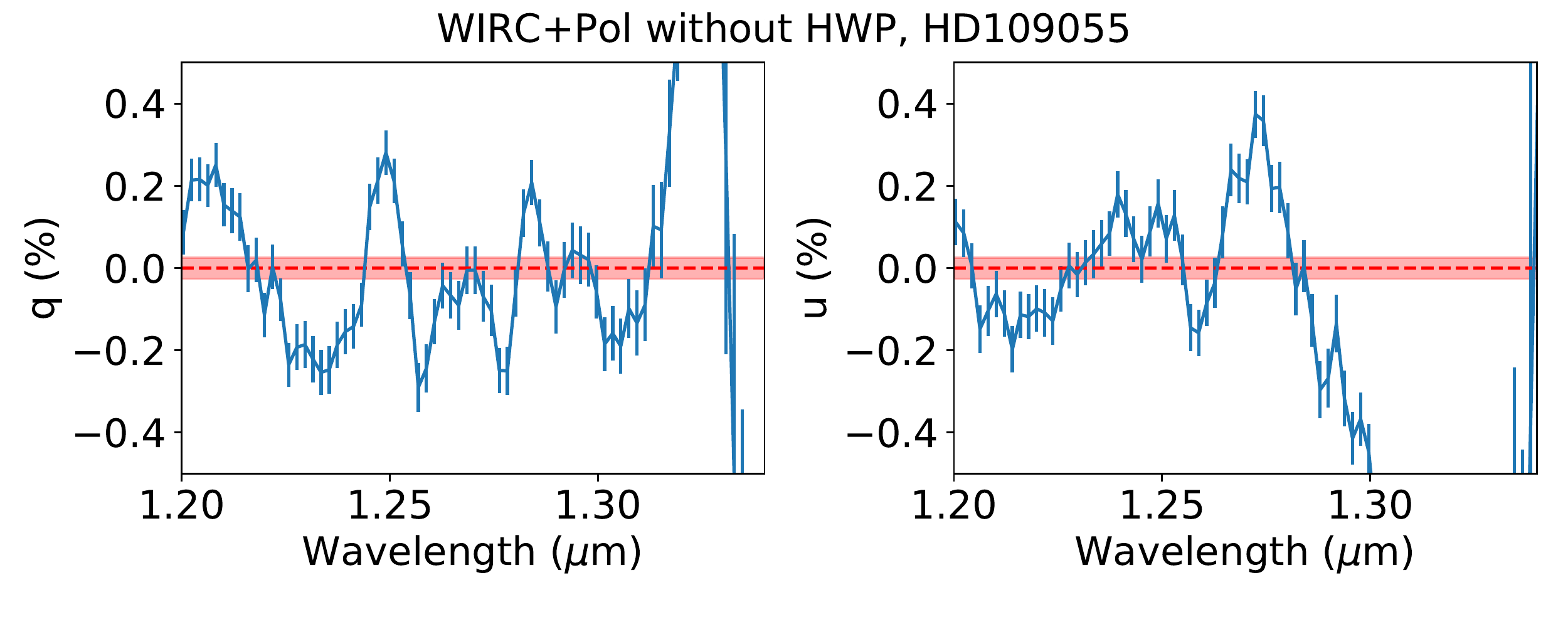} \\
    \includegraphics[width=0.8\textwidth]{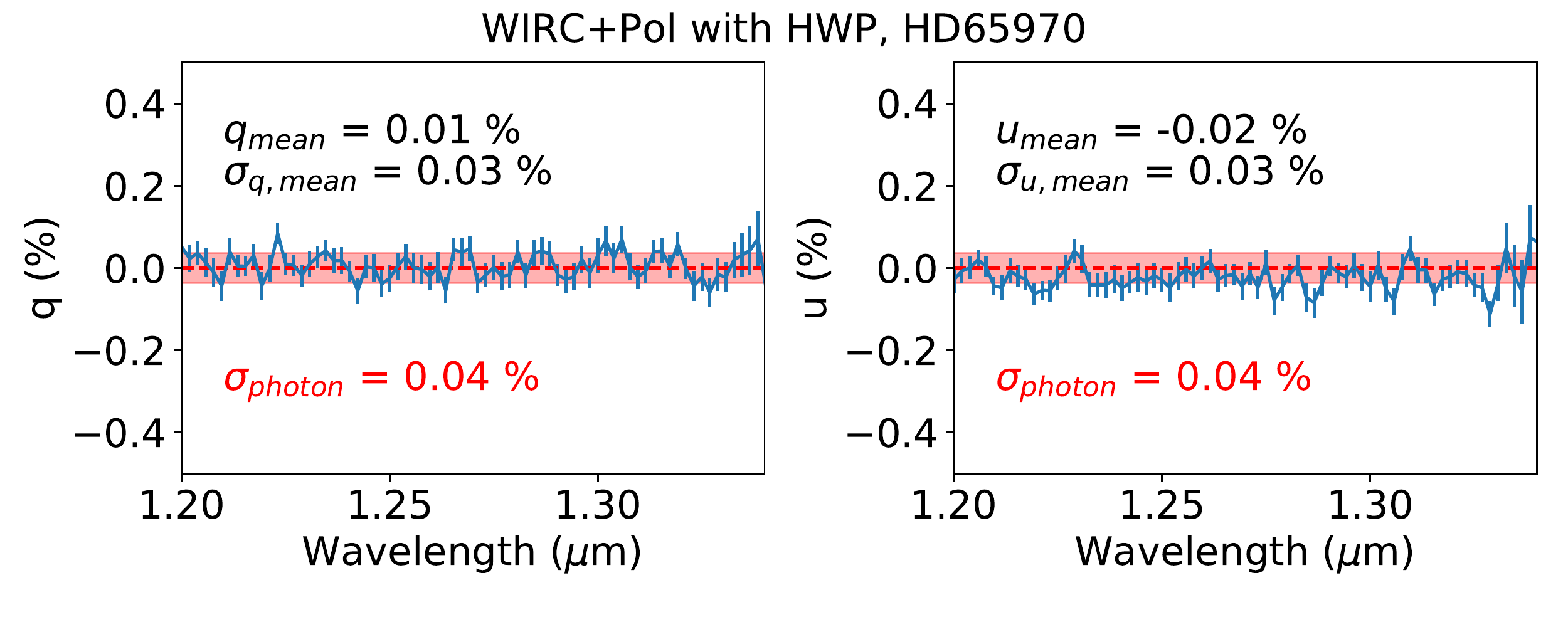}
    \caption{Top: normalized Stokes parameters $q$ and $u$ in percent of an unpolarized standard star HD\,109055 observed with WIRC+Pol without the HWP. The dashed line at 0\% indicates the expectation value, while the shaded region shows the expected uncertainties from photon noise. 
    Bottom: the same of HD\,65970 observed with WIRC+Pol with the HWP. The mean $q$ and $u$ and the associated uncertainties in the wavelength band shown are annotated on the plots. In this case, the measured $q$ and $u$ are within 1$\sigma$ from zero, and the uncertainties are consistent with the photon noise, shown as the shaded region.}
    \label{fig:qu_compare}
\end{figure}

Fig.~\ref{fig:qu_compare} compares the measured $q$ and $u$ of an unpolarized star HD\,65970 (J = 7.689) observed for 200 s using the HWP to that of an unpolarized star HD\,109055 (J = 8.731) observed for 605 s without the HWP.
The latter is the same observation presented in Fig.~6 top left in Ref.~\citenum{tinyanont2019} as the best zero polarization measurement with the old WIRC+Pol.
The dashed lines mark zero polarization, expected for these stars, and the band around them indicate uncertainties expected from photon noise. 
The measurements are plotted with error bars computed from several sets of observations obtained for each source.
It is evident that the strong, wavelength dependent instrumental systematics, clearly visible in old WIRC+Pol data without the HWP, have been eliminated. 
The measurements for HD\,65970 with the HWP are zero to within 1$\sigma$, and the uncertainties derived from the data are the same as what expected from photon noise.

\subsection{Faint source, slitless}
In addition to a bright unpolarized standard stars, we also observed several brown dwarfs (BD) and placed strict polarization upper limits on some of them. 
These observations were part of the BD and supernovae (SNe) spectropolarimetric surveys being conducted with WIRC+Pol. 
Fig.~\ref{fig:faint_qu} (top) shows the normalized Stokes parameters $q$ and $u$ for a BD, 2MASS J18071593+5015316. 
The BD's magnitude was $J = 12.934$, and it was observed outside the slit for 80 min on 2019 July 16.
The measured polarization was consistent with zero, and we noted that the uncertainties derived from the data (plotted as error bars) were 0.1\%, consistent with the photon noise, which is marked by the red band on the plots. 
This result demonstrated that the background subtraction in the slitless mode was not biasing the measurement, even when the observations were background dominated. 

\subsection{Faint source, in-slit}
\begin{figure}
    \centering
    \includegraphics[width=0.8\textwidth]{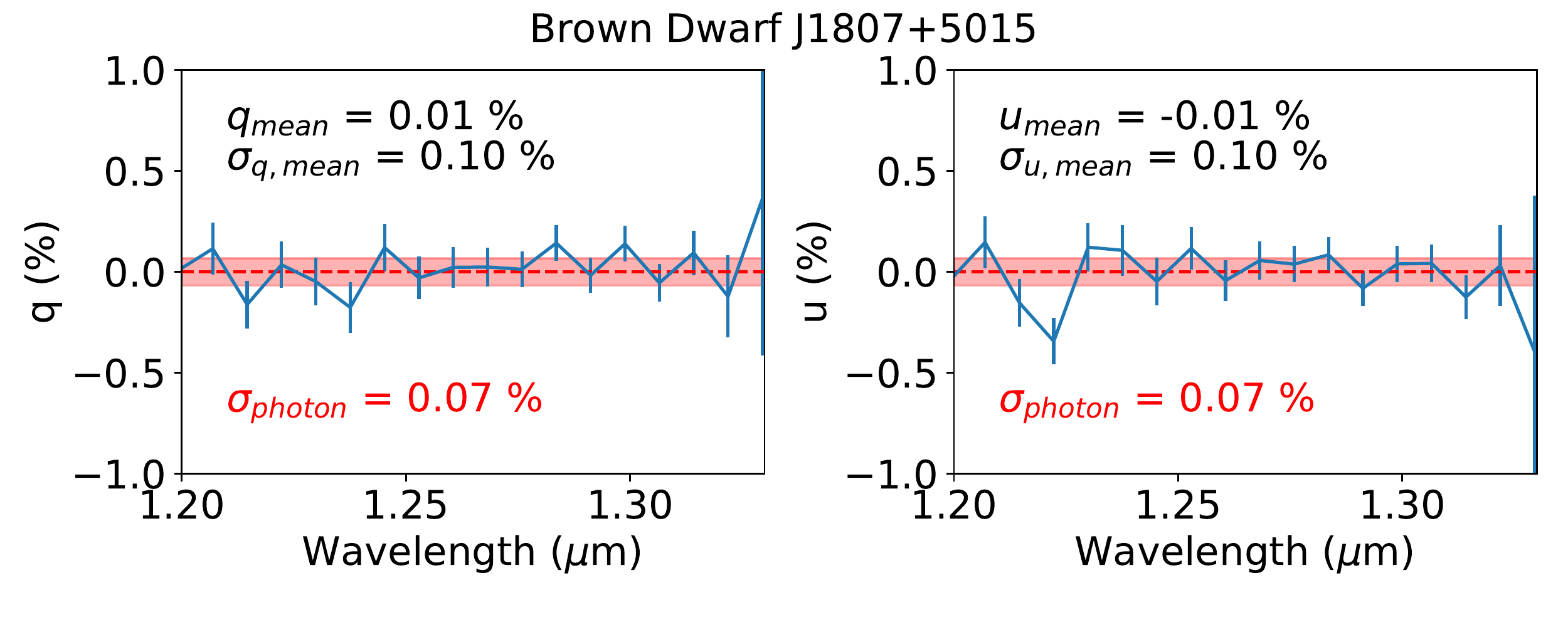} \\
    \includegraphics[width=0.8\textwidth]{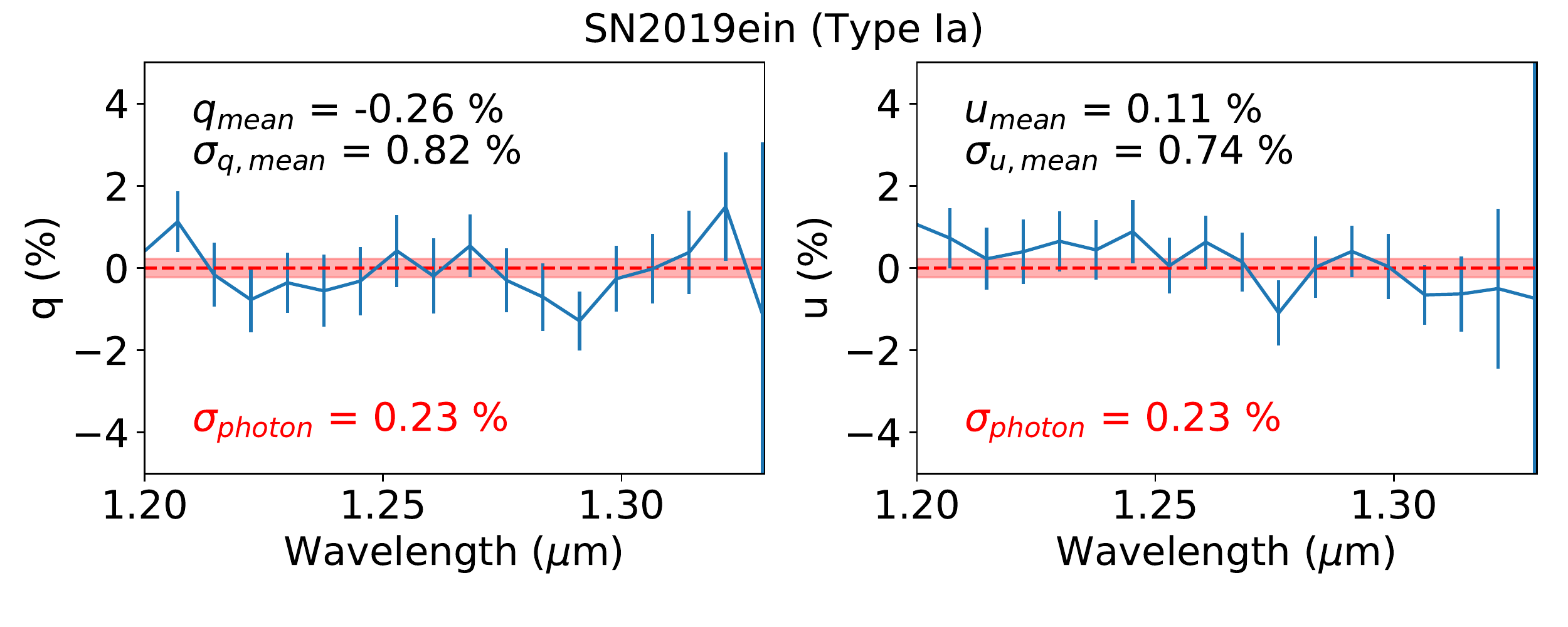}
    \caption{Top: Normalized Stokes parameters $q$ (left) and $u$ (right) of a brown dwarf J1807+5015, $J = 12.934$, observed for 80 min.
    The dashed lines mark zero polarization and the red bands are the expected photon noise.
    The brown dwarf was observed outside the slit.
    Bottom: Same for SN\,2019ein, a thermonuclear Type Ia SN, which typically has zero polarization at this phase. The SN was at $J\approx 14.5$, and it was observed in the slit for 84 min. }
    \label{fig:faint_qu}
\end{figure}

Lastly, Fig.~\ref{fig:faint_qu} (bottom) shows the same plot for SN\,2019ein, a Type Ia SN, observed at 14 days post-discovery\cite{tonry2019,burke2019}.
The source was at $J \approx 14.5$ and the total exposure time was 84 min.
At this phase, a typical SN Ia is not expected to be polarized\cite{wang2008}, and our observations did not show significant departure from null polarization. 
We note that, unlike for the slitless observations, the uncertainties, which are the standard error of the mean derived from the series of exposures, are about a factor of four larger than what expected from photon noise. 
The cause of this was most likely the variable slit loss, resulting from the source moving slightly inside the slit between HWP rotations. 
This also demonstrates that the slit-induced polarization discussed in the prior section does not affect our measurements.

\section{Future characterizations}\label{sec:mueller}
\subsection{Angle of polarization zero point}
The HWP installed in WIRC+Pol provides half-wave retardation over a large wavelength range but the fast axis is not wavelength independent. 
The angle of polarization zero point as a function of wavelength needs to be determined by observing both stars with known polarization and unpolarized standard stars with the LP. 
This measurement could be done with existing data. 

\subsection{The Mueller Matrix}
The instrument's Mueller matrix, with circular polarization terms omitted, is the matrix in this following equation 
\begin{equation}
    \begin{bmatrix} 
1 \\
q_{\rm obs} \\
u_{\rm obs} 
\end{bmatrix} = 
    \begin{bmatrix} 
1 & 0 & 0 \\
q_{\rm IP}& \eta_q & \chi_{uq} \\
u_{\rm IP} & \chi{qu} & \eta_{u}
\end{bmatrix}     \begin{bmatrix} 
1 \\
q \\
u 
\end{bmatrix}
\end{equation}
It relates the intrinsic normalized Stokes vector to the observed one.

With the observations of unpolarized sources shown in the last section, we have already demonstrated that if $q = u = 0$, then the observed polarizations are also $q_{\rm obs} = u_{\rm obs} = 0$ to within the measured uncertainties. 
Hence, the upper limits of the instrumental polarization terms are $q_{\rm IP}, u_{\rm IP} < 0.03\%$. 
To measure the other terms in the Mueller matrix, namely the polarization efficiencies ($\eta$) and cross-talks ($\chi$), we need observations of polarized standard stars with known, non-zero $q$ and $u$, as functions of wavelength in the literature. 
The issue is that spectropolarimetric measurements of standard stars in the near-IR, unlike the broadband measurements, are rare.  
Therefore, we plan to use the combination of (i) Serkowski's law fitted to broadband measurements presented in Ref.~\citenum{whittet1992} and (ii) unpolarized standard stars observed with the LP in the beam. 
These observations have been obtained and we are reducing the data to measure the remaining terms in the Mueller matrix of WIRC+Pol and determine whether they are constant or are functions of telescope pointing. 

\section{Conclusion}\label{sec:conclusion}
WIRC+Pol is a low-resolution ($R\approx 100$), near-infrared (1.1-1.8 $\mu$m, \textit{JH} bands) spectropolarimeter. 
The instrument employs a liquid crystal polymer-based PG to perform polarimetric beam splitting and spectral dispersion simultaneously.
The instrument produces four spectral traces on the detector, probing 0$^\circ$, 90$^\circ$, 45$^\circ$, and 135$^\circ$ angles of polarization at the same time, allowing it to measure Stokes parameters $q$ and $u$ in one shot.
We found from commissioning data from the instrument in 2017-2018 that the polarimetric systematics from comparing fluxes in different traces across the detector were significant (few \%) and were slowly changing with time.
Thus, they were difficult to remove.
In March 2019, a HWP and a LP were installed in front of the instrument, allowing us to modulate the angle of polarization probed by each of the four spectral traces.
As a result, the Stokes parameters could be computed using the flux ratio method described earlier, which could measure and remove systematics due to non-common path and atmospheric effects. 
We verified the performance of the HWP by showing that the systematics were indeed removed, both for bright standard stars and faint sources. 
The polarimetric uncertainties for slitless observations now seem to be only photon noise limited. 
In addition to the HWP/LP, a new focal plane mask with a slit manufactured with EBL was also installed, allowing WIRC+Pol to more easily observe faint targets with reduced background.
We are developing efficient observational guidelines for slit observations to reduce overhead from target acquisition and dithering. 
For observations in slit, the polarimetric uncertainties are a factor of a few larger than the photon noise, likely from variable slit loss.
Future instrumental calibrations include (i) measuring the angle of polarization zero point as a function of wavelength and (ii) measuring the polarimetric efficiency and cross-talk, the remaining possibly non-zero terms of the Mueller matrix of WIRC+Pol.
Scientific observations with WIRC+Pol has commenced with a large brown dwarf spectropolarimetric survey and a nearby supernovae survey ongoing. 
First results from these surveys are expected later this year.

\label{sec:intro}  

\bibliography{report} 

\newcommand{\noop}[1]{}
\begin{thebibliography}{10}

\bibitem{manchado2004}
{Manchado}, A., {Barreto}, M., {Acosta-Pulido}, J., {Ballesteros}, E.,
  {Barreto}, R., {Cadavid}, E., {Carrillo}, J., {Charcos}, M., {Correa}, S.,
  {Delgado}, J.~M., {Dominguez-Tagle}, C., {Gonzalez}, O., {Hernandez}, E.,
  {Lopez}, R., {Moreno}, H., {Olives}, J., {Peraza}, L., {Prada}, F.,
  {Redondo}, P., {Sanchez}, V., {Sosa}, N., {Tenegi}, F., and {Vidal}, M.~J.,
  ``{First light for LIRIS (long-slit intermediate-resolution infrared
  spectrograph)},'' in [{\em Ground-based Instrumentation for
  Astronomy}{\nolinebreak\hspace{0.1em}]},  {Moorwood}, A.~F.~M. and {Iye}, M.,
  eds., {\em \procspie} {\bf 5492},  1094--1104 (Sept. 2004).

\bibitem{watanabe2018}
{Watanabe}, M., {Pyo}, T.-S., {Terada}, H., {Hattori}, T., {Hayano}, Y.,
  {Minowa}, Y., {Oya}, S., {Hattori}, M., {Kudo}, T., {Morii}, M., {Hashimoto},
  J., and {Tamura}, M., ``{Near-infrared adaptive optics imaging- and
  spectro-polarimetry with the infrared camera and spectrograph of the Subaru
  Telescope},'' in [{\em \procspie}{\nolinebreak\hspace{0.1em}]},  {\em Society
  of Photo-Optical Instrumentation Engineers (SPIE) Conference Series} {\bf
  10702},  107023V (Jul 2018).

\bibitem{tokunaga1998}
{Tokunaga}, A.~T., {Kobayashi}, N., {Bell}, J., {Ching}, G.~K., {Hodapp},
  K.-W., {Hora}, J.~L., {Neill}, D., {Onaka}, P.~M., {Rayner}, J.~T.,
  {Robertson}, L., {Warren}, D.~W., {Weber}, M., and {Young}, T.~T.,
  ``{Infrared camera and spectrograph for the SUBARU Telescope},'' in [{\em
  \procspie}{\nolinebreak\hspace{0.1em}]},  {Fowler}, A.~M., ed., {\em Society
  of Photo-Optical Instrumentation Engineers (SPIE) Conference Series} {\bf
  3354},  512--524 (Aug 1998).

\bibitem{donati2018}
{Donati}, J.-F., {Kouach}, D., {Lacombe}, M., {Baratchart}, S., {Doyon}, R.,
  {Delfosse}, X., {Artigau}, {\'E}., {Moutou}, C., {H{\'e}brard}, G., {Bouchy},
  F., {Bouvier}, J., {Alencar}, S., {Saddlemyer}, L., {Par{\`e}s}, L., {Rabou},
  P., {Micheau}, Y., {Dolon}, F., {Barrick}, G., {Hernandez}, O., {Wang},
  S.~Y., {Reshetov}, V., {Striebig}, N., {Challita}, Z., {Carmona}, A.,
  {Tibault}, S., {Martioli}, E., {Figueira}, P., {Boisse}, I., and {Pepe}, F.,
  [{\em {SPIRou: A NIR Spectropolarimeter/High-Precision Velocimeter for the
  CFHT}}{\nolinebreak\hspace{0.1em}]},  107 (2018).

\bibitem{millar2014}
{Millar-Blanchaer}, M., {Moon}, D.-S., {Graham}, J.~R., and {Escuti}, M.,
  ``{Polarization gratings for visible and near-infrared astronomy},'' in [{\em
  Advances in Optical and Mechanical Technologies for Telescopes and
  Instrumentation}{\nolinebreak\hspace{0.1em}]},  {\em \procspie} {\bf 9151},
  91514I (July 2014).

\bibitem{packham2010}
{Packham}, C., {Escuti}, M., {Ginn}, J., {Oh}, C., {Quijano}, I., and
  {Boreman}, G., ``{Polarization Gratings: A Novel Polarimetric Component for
  Astronomical Instruments},'' {\em \pasp}~{\bf 122},  1471--1482 (Dec. 2010).

\bibitem{escuti2006}
Michael J.~Escuti, Chulwoo~Oh, C. S. C. B. D. J.~B., ``Simplified
  spectropolarimetry using reactive mesogen polarization gratings,'' {\em
  Proc.SPIE}~{\bf 6302},  6302 -- 6302 -- 11 (2006).

\bibitem{wilson2003}
{Wilson}, J.~C., {Eikenberry}, S.~S., {Henderson}, C.~P., {Hayward}, T.~L.,
  {Carson}, J.~C., {Pirger}, B., {Barry}, D.~J., {Brandl}, B.~R., {Houck},
  J.~R., {Fitzgerald}, G.~J., and {Stolberg}, T.~M., ``{A Wide-Field Infrared
  Camera for the Palomar 200-inch Telescope},'' in [{\em Instrument Design and
  Performance for Optical/Infrared Ground-based
  Telescopes}{\nolinebreak\hspace{0.1em}]},  {Iye}, M. and {Moorwood},
  A.~F.~M., eds., {\em \procspie} {\bf 4841},  451--458 (Mar. 2003).

\bibitem{tinyanont2019}
{Tinyanont}, S., {Millar-Blanchaer}, M.~A., {Nilsson}, R., {Mawet}, D.,
  {Knutson}, H., {Kataria}, T., {Vasisht}, G., {Henderson}, C., {Matthews}, K.,
  {Serabyn}, E., {Milburn}, J.~W., {Hale}, D., {Smith}, R., {Vissapragada}, S.,
  {Santos}, Louis~D., J., {Kekas}, J., and {Escuti}, M.~J., ``{WIRC+Pol: A
  Low-resolution Near-infrared Spectropolarimeter},'' {\em \pasp}~{\bf 131},
  025001 (Feb 2019).

\bibitem{wiktorowicz2009}
{Wiktorowicz}, S.~J., {\em {Unambiguous black hole mass from polarimetry and
  application to hot Jupiters}}, PhD thesis, California Institute of Technology
  (Jan 2009).

\bibitem{zapatero2011}
{Zapatero Osorio}, M.~R., {B{\'e}jar}, V.~J.~S., {Goldman}, B., {Caballero},
  J.~A., {Rebolo}, R., {Acosta-Pulido}, J.~A., {Manchado}, A., and {Pe{\~n}a
  Ram{\'{\i}}rez}, K., ``{Near-infrared Linear Polarization of Ultracool
  Dwarfs},'' {\em \apj}~{\bf 740},  4 (Oct. 2011).

\bibitem{slater1959}
Slater, J.,  [{\em Microwave transmission}{\nolinebreak\hspace{0.1em}]}, Dover
  book ; S564, Dover Publications (1959).

\bibitem{ismail1986}
{Ismail}, M.~A., ``{Transmission of Light by Slit Designed for Astronomical
  Spectrograph},'' {\em \apss}~{\bf 122},  1--32 (May 1986).

\bibitem{heiles2000}
{Heiles}, C., ``{9286 Stars: An Agglomeration of Stellar Polarization
  Catalogs},'' {\em \aj}~{\bf 119},  923--927 (Feb. 2000).

\bibitem{tonry2019}
{Tonry}, J., {Denneau}, L., {Heinze}, A., {Weiland}, H., {Flewelling}, H.,
  {Stalder}, B., {Rest}, A., {Stubbs}, C., {Smith}, K.~W., {Smartt}, S.~J.,
  {Young}, D.~R., {McBrien}, O., {O'Neill}, D., {Clark}, P., {Fulton}, M.,
  {McCormack}, A., and {Wright}, D.~E., ``{ATLAS Transient Discovery Report for
  2019-05-01},'' {\em Transient Name Server Discovery Report}~{\bf 2019-678},
  1 (May 2019).

\bibitem{burke2019}
{Burke}, J., {Arcavi}, I., {Howell}, D.~A., {Hiramatsu}, D., {McCully}, C., and
  {Valenti}, S., ``{Global SN Project Transient Classification Report for
  2019-05-03},'' {\em Transient Name Server Classification Report}~{\bf
  2019-701},  1 (May 2019).

\bibitem{wang2008}
{Wang}, L. and {Wheeler}, J.~C., ``{Spectropolarimetry of supernovae.},'' {\em
  \araa}~{\bf 46},  433--474 (Sep 2008).

\bibitem{whittet1992}
{Whittet}, D.~C.~B., {Martin}, P.~G., {Hough}, J.~H., {Rouse}, M.~F., {Bailey},
  J.~A., and {Axon}, D.~J., ``{Systematic variations in the wavelength
  dependence of interstellar linear polarization},'' {\em \apj}~{\bf 386},
  562--577 (Feb. 1992).

\end{thebibliography}
\bibliographystyle{spiebib} 

\end{document}